# Far-Field Super-Resolution Imaging By Nonlinear Excited Evanescent Waves

Zhihao Zhou[1], Wei Liu[2], Jiajing He[1], Lei Chen[2], Xin Luo[1], Dongyi Shen[2],

Jianjun Cao[3], Yaping Dan[1], Xianfeng Chen[2] and Wenjie Wan[1,2*]

[1]The State Key Laboratory of Advanced Optical Communication Systems and Networks,

University of Michigan-Shanghai Jiao Tong University Joint Institute,

Shanghai Jiao Tong University, Shanghai 200240, China

[2]MOE Key Laboratory for Laser Plasmas and Collaborative Innovation Center of IFSA,

School of Physics and Astronomy, Shanghai Jiao Tong University, Shanghai 200240, China

[3]School of Science, Jiangnan University, Wuxi 214122, China

* Corresponding author: Wenjie Wan wenjie.wan@sjtu.edu.cn

**Abbe's resolution limit, one of the best-known physical limitations, poses a great challenge for any wave systems in imaging, wave transport, and dynamics. Originally formulated in linear optics, this Abbe's limit can be broken using nonlinear optical interactions. Here we extend the Abbe theory into a nonlinear regime and experimentally demonstrate a far-field, label-free, and scan-free super-resolution imaging technique based on nonlinear four-wave mixing to retrieve near-field scattered evanescent waves, achieving sub-wavelength resolution of $\lambda/15.6$. This method paves the way for application in biomedical imaging, semiconductor metrology, and photolithography.**

## 1. Introduction

The spatial resolution of an imaging system is limited by the Abbe theory[1], posing a great challenge for many areas like biomedical imaging, astronomy, and photolithography. For example, direct live images are crucial to understanding biological processes at the subcellular level, e.g. virus behaviors[2,3]; the finest nanostructures fabricated by photolithography on semiconductor chips is also defined by the Abbe's limit. According to Abbe's theory, sub-wavelength image features are usually associated with near-field evanescent waves, which decay exponentially with distance similar to electron wavefunctions in quantum tunneling[4] and diffusive waves in Anderson localization[5,6]. Insufficient far-field detection of these evanescent fields in conventional optics with finite numerical apertures ultimately precludes imaging resolution better than $\lambda/2$. Like scanning tunneling electron technique in condensed-matter physics converting tunneling currents into conducting ones[7], near-field scanning optical microscope (NSOM) can improve the resolution by converting evanescent waves into propagating ones[2] but requires near-field scanning. Under the same framework, recent advancements in metamaterials allow similar conversions using superlens[8], hyperlens[9,10], and surface plasmon polaritons (SPP)[11] to achieve super-resolution imaging but with demanding nanofabrications. On the other hand, X-ray and electron microscopes, even with nanometer resolution, may be potentially harmful to biomedical applications, it is still highly desirable to break the Abbe's limit and realize super-resolution imaging for optical waves.

Nonlinear optics may offer an alternative way to beat the Abbe's limit, which is originally formulated for linear waves[12–14]. By introducing nonlinear spatial wave mixing into an imaging system, it becomes possible to retrieve those undetected waves in the far-field and reconstruct to improve the resolution[13], effectively, generalizing the Abbe theory into nonlinear optics regime. However, evanescent waves are absent in prior works, the imaging resolutions have not reached the wavelength scale yet till this work. Meanwhile, evanescent waves can be manipulated through nonlinear wave mixings[14–16] based on surface phase-matching conditions, e.g. free-space coupling of SPP[17,18], dark-field imaging[19]. These works show a unique way to nonlinearly couple non-propagating evanescent waves with propagating ones into far-field, addressing the key issue for the aforementioned super-resolution imaging problem in the framework of nonlinear Abbe theory. Moreover, it may enable a fluorophore-label-free imaging method in contrast to the existing techniques like stimulated emission depletion (STED) microscopy[20], stochastic optical reconstruction microscopy (STORM)[3], structured illumination microscopy (SIM)[21,22], where label-free imaging techniques are highly desirable not only in biomedical imaging[23], but also in other areas like semiconductor metrology processes.

In this work, we experimentally show nonlinear wave mixings including evanescent waves in the framework of nonlinear Abbe theory and demonstrate a far-field, label-free, and scan-free super-resolution imaging scheme to resolve sub-wavelength structures on a semiconductor silicon-on-insulator (SOI) wafer. To break the Abbe's resolution limit, a nonlinear four-wave mixing (FWM) technique is implemented to excite localized near-field evanescent wave-based illumination with large spatial wave vectors, such that near-field waves containing the finest sub-wavelength imaging features can be converted into propagating ones for far-field detections, effectively enlarging the numerical apertures. Combined with an iterative Fourier ptychography method[24,25], the reconstructed images can reach a resolution limit down to $\lambda/15.6$ with respect to the input probe's wavelength. Moreover, this FWM imaging scheme can also cooperate with nano-slit grating structures, which theoretically can provide additional resolution enhancement. This technique may offer a new way for critical imaging tasks sensitive to fluorescent labels like biomedical applications, semiconductor processes as well as photolithography.

## 2. Methods

According to linear Abbe's theory, a sub-wavelength object illuminated by a coherent wave with wave vector $\mathbf{k}$ can cause scattering waves to radiate over a wide-angle into the far-field. Meanwhile, a portion of scattered light with large wave vectors becomes evanescent and confined only to the object's surface in the near field as shown in Fig. 1(a). Such evanescent components correspond to large $k$ portions in the Fourier space [Fig. 1(b)], determining the finest feature of the object. How to retrieve these near-field evanescent waves in the far-field is the key to tackle the problem of Abbe's diffraction limit[26]. Linearly, evanescent waves can be scattered off by a sharp sub-wavelength tip in the case of NSOM and convert into propagating waves for far-field detection. Similarly, we can expect that nonlinear wave mixing under certain phase-matching conditions allows similar conversions between near-field evanescent waves and far-field incident waves as shown in prior works[17,18,27]. Here we purposely excite a localized evanescent wave with wave vector $\mathbf{k}_{3,eva}$ through nonlinear FWM with far-field launched pump and probe beams [Fig. 1(a)]. Consequently, the excited evanescent waves will be scattered off by the sub-wavelength target into radiating ones at various angles.

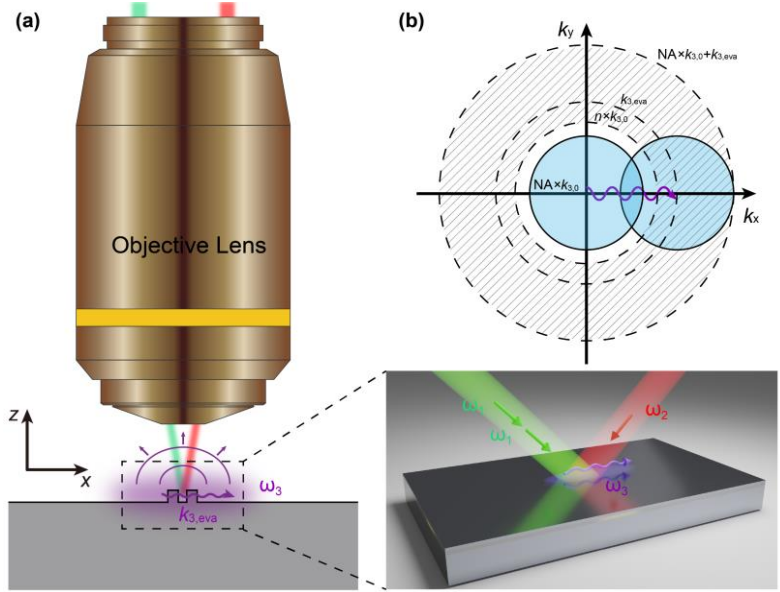

**FIG. 1. The working principle of far-field super-resolution imaging based on nonlinear excited evanescent waves** (a) Schematic of the far-field super-resolution imaging process with localized evanescent wave illumination generated by FWM process at the interface. The zoomed inset shows the FWM process takes place on nonlinear medium's interface (b) Basic mechanism illustrated in Fourier space. NA and $k_{3,0}$ represent numerical aperture, the free-space wavenumber of the FWM signal beam, respectively. $NA \times k_{3,0}$ determines the cut-off frequency of the imaging system; correspondingly, the blue-circled area represents the system's passband. Striped-shadow regions represent evanescent fields, which carry sub-wavelength details of the imaging target. In our method, we introduce the evanescent wave vector $k_{3,eva}$ as illumination light, which can map evanescent fields into propagating ones. The final accessible region in Fourier space can be extended to the utmost dashed circle and effectively enlarges the NA.

In the spatial-frequency domain, i.e. the k-space [Fig. 1(b)], this scattering process can be described by in-plane momentum conservation[28]:

$$|k_{3,x}| = |k_{3,eva} - 2\pi/\Lambda| < n_3 k_{3,0} \qquad (1)$$

where $k_{3,0}$ is the wavenumber of FWM in a vacuum, $\Lambda$ represent various spatial features of the target. In reality, the imaging target contains a wide spectrum of spatial modes, extending to a wide-spreading disk in the k-space [Fig. 1(b)]. Under a normal-incident illumination, these spatial modes collected in the far-field through an imaging lens with limited NA lie within the range of $|2\pi/\Lambda| < NA \times k_{3,0}$ in the k-space. In contrast, if illuminated by the evanescent wave excited by FWM, the effective passband in the k-space of the same imaging system will be shifted by $\mathbf{k}_{3,eva}$ [Fig. 1(b)], meaning that most parts of the passband lie within the evanescent regime of the target spectrum, where lie the finest sub-wavelength features of imaging object. In this manner, these evanescent waves with large wave vectors carrying sub-wavelength details can be retrieved in the far-field with a conventional imaging objective lens. Later, combining with certain numerical reconstruction methods, we shall be able to restore far-field super-resolved images with subwavelength resolution. In a similar approach, coherent total internal reflection (TIR) microscopy relied on linear TIR excited evanescent waves[29] can also enhance the imaging resolution, but limited by medium refractive indexes.

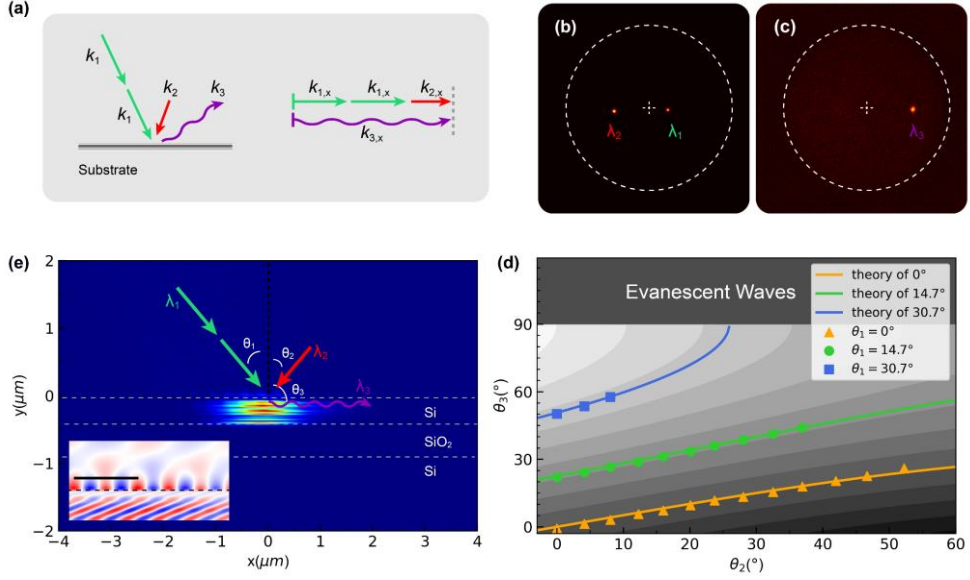

**FIG. 2. Demonstration of wave vector control in surface FWMs and local excitation of evanescent waves** (a) Illustration of the FWM process takes place at the interface and the partial phase-matching condition. (b-c) Fourier space images of reflected pump/probe beams $\lambda_1$, $\lambda_2$ and signal beam $\lambda_3$ taken by EMCCD, characterizing incident angles and output angle, respectively. (d) Dependence of FWM output angle $\theta_3$ on probe incident angle $\theta_2$ under different pump angles $\theta_1$. (e) Numerical simulation result of FWM signal field distribution, which is the case of an evanescent wave with large transverse wave vector $\mathbf{k}_{3,\text{eva}}$ localized at the top film of SOI. The inset shows the signal's amplitude variation along with the interface, where the scale bar represents the wavelength of FWM.

The key mechanism to excite near-field evanescent waves through FWM can be understood from nonlinear phase-matching conditions at the interface. In linear optics, the laws of reflection and refraction at an interface can be directly deduced from electromagnetic boundary conditions laid out by Fresnel formalism[30]. Under the same framework, the well-known linear Snell's law can be generalized into a nonlinear regime where light waves are incident onto a nonlinear medium boundary by satisfying boundary conditions for all individual frequencies during nonlinear conversions[31]. For the case of FWMs, the resulting equations governing reflection and refraction have been developed in Ref[15,27] as well as our prior work[14], where the nonlinear reflection law reads:

$$k_3 \sin(\theta_3) = 2k_1 \sin(\theta_1) - k_2 \sin(\theta_2) \qquad (2)$$

here $k_i = n_i \omega_i / c$ ($i = 1, 2, 3$) are wave vectors of the pump, probe, and FWM signal beams, respectively; $n_i$ are refractive indexes of the surrounding medium, $\theta_i$ represent their incident/output angles. Effectively, the relations of these transverse wave vectors represent in-plane momentum conservation law as shown in Fig. 2(a). Such a partial phase-matching condition only manifests itself near surfaces[19,32] and can be enhanced by thin-film structures[27]. More importantly, this reduced-dimension requirement of phase matching can lead to intriguing results where the transverse component of FWM wave $k_{3,x}$ (denoted by $k_{3,\text{eva}}$ for evanescent waves later) can exceed its free-space magnitude $k_3$, i.e. propagating waves. As a result, its corresponding vertical component becomes pure imaginary according to $k_{3,z} = i\sqrt{k_{3,x}^2 - k_3^2}$, making the wave evanescent. These generated evanescent waves are bounded at the interface, but with large in-plane wave vector, which is crucial for the aforementioned scattering imaging process.

Experimentally, we first verify Eq. (2) in a reflection configuration by synchronously launching a pump and a probe beam onto a flat SOI wafer to excite near-field FWM as shown in Fig. 2(a). Here two pump photons at $\lambda_1=532\text{nm}$ from a frequency-doubled fiber femtosecond laser and a probe photon at $\lambda_2=780\text{nm}$ from a home-built optical parametric oscillator (OPO) interact together, exciting one FWM signal photon at $\lambda_3=403\text{nm}$ (details in supplementary materials and Ref[33]). Both the incident and output beams are collected at the back focal plane [Fig. 2(b), 2(c)], which vividly reveal the wave vector distributions in Fourier space. The measured reflection angles of FWM waves $\theta_3$ exhibit an upward trend [Fig. 2(d)] with increasing incident angles of the probe $\theta_2$ for a given angle $\theta_1$, well fitted to the theoretical curves predicted by Eq. (2). Furthermore, this reflection angle $\theta_3$ can exceed 90°, making the FWM evanescent. For example, for the incident angle $\theta_1=30.7°$ (blue curve), the FWM waves become evanescent above the angle $\theta_2=25.9°$ (the maximum collected angle of FWM is limited to ~58.9° due to the NA), denoted by a gray region in Fig. 3(d), which is the region of interest to explore our evanescent wave-based sub-wavelength imaging. Also, such nonlinear excited evanescent waves have been previously demonstrated for surface plasmon coupling[17,18], dark field imaging[19].

To illustrate the case of evanescent wave excitation, we perform a numerical simulation of surface FWM by finite-difference time-domain (FDTD) method. The calculated $|\mathbf{E}|^2$ distribution of the FWM signal shows that the signal wave is localized near the focal spot inside the top silicon layer of the SOI wafer [Fig. 2(e)]. This is a sharp contrast to nonlinear excited surface plasmon mode near a metal-dielectric interface[11,18], where surface plasmons can further propagate under certain phase matching. Such nonlinear excited FWM waves serve as localized evanescent light sources for sub-wavelength imaging purposes later. According to the aforementioned discussions on imaging resolution, it is essential to obtain large transverse wave vectors for excited evanescent waves. As shown in the inset of Fig. 2(e), the effective wavelength of such evanescent wave has been reduced to ~200nm, half of its free-space wavelength (~403nm), leading to the relation of wave vectors as $k_{3,\text{eva}}=2k_{3,0}$, which is the same result according to Eq. (2) for given input angles in the simulation ($\theta_1,\theta_2=40°$). Effectively, by varying the input angles of pump and probe beams, we manage to locally excite evanescent waves with variable wave vectors. Such FWM processes enable an active, flexible manner to control signal beam's wave vectors for imaging processes later.

## 3. Results

To implement an evanescent wave-based sub-wavelength imaging, we compare imaging qualities under several illumination schemes in Fig. 3. Consider a double-slit with a width of $w$ and center-to-center distance $a$ [Fig. 3(j)] illuminated by an evanescent wave with wave vector $k_{3,\text{eva}}$, the reflected far-field image through an NA limited imaging system in the spatial domain can be formulated as[34]:

$$I(x)=|s(x)+\exp(ik_{3,\text{eva}}a)s(x-a)|^2 \qquad (3)$$

where $s(x)=h(x)*[\exp(ik_{3,\text{eva}}x)\times\text{rect}(\frac{x}{w})]$ is the amplitude distribution of a single slit function centered at $x=0$, i.e. $\text{rect}(\frac{x}{w})$, and $h(x)$ is point spread function (PSF) of the imaging system. The extra phase term induced by the illumination beam $\exp(ik_{3,\text{eva}}x)$ effectively shifts the angular spectrum of the images in k-space by $k_{3,\text{eva}}$ as shown in Fig. 3(b), which represents the finer portion of the image with better resolution. For example, the relative phase term $\exp(ik_{3,\text{eva}}a)$ between the first $s(x)$ and the second slit $s(x-a)$ can be out-of-phase in the regime of $\frac{\pi}{2} < k_{3,\text{eva}} a < \frac{3\pi}{2}$, forming a dip in the far-field image [Fig. 3(e), 3(h)]. In contrast, the same feature can not be resolved under normal illumination when $k_{2,x} = 0$ [Fig. 3(d), 3(g)].

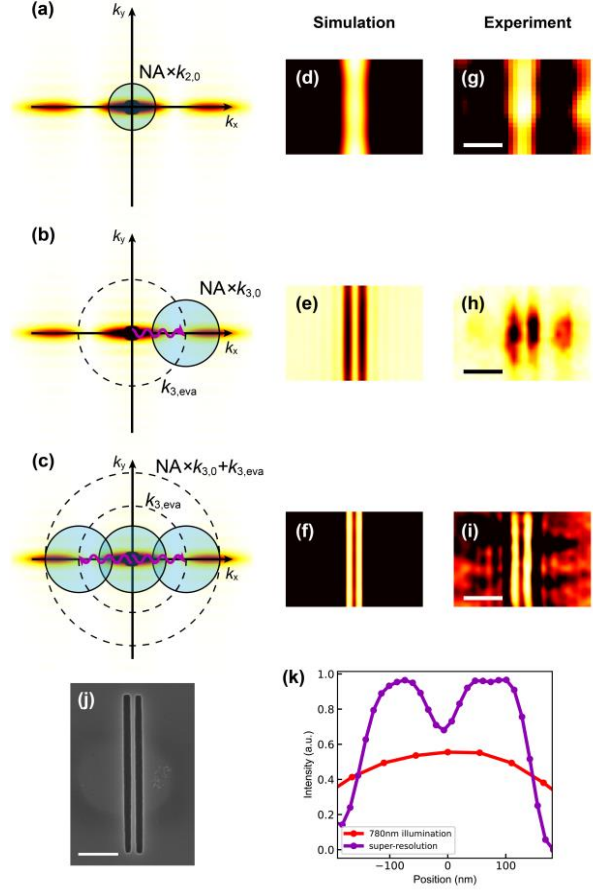

**FIG. 3. Demonstration of super-resolution imaging using FWM evanescent wave illumination.** (a)-(c) Fourier space representation of (a) probe beam at 780nm with normal illumination (b) FWM signal at 403nm with $k_{3,\text{eva}}$ illumination (c) the complete imaging method, respectively. (d-f) Simulated images of a pair of 90nm-wide slits spaced 50nm apart, corresponding to cases of (a)-(c). (g-i) Experimental results. Scale bar: 500nm (j) Scanning electron microscopy (SEM) image of the two-slit target (k) Inverse-intensity cross-section comparison of probe beam normal illumination image and proposed super-resolution method.

Previously, a similar idea has been applied in total internal reflection microscopy (TIR) to improve the resolution[29]. But the key problem is to retrieve images and improve the corresponding resolution under such evanescent wave illumination. Here we restore sub-wavelength images using a computational imaging technique named Fourier ptychography (FP)[24,25], by stacking multiple low-resolution images with pre-determined illumination angles and iteratively reconstructing a final image with much-enhanced resolution (see supplementary materials). Previously this technique has shown great successes in wide-field, high-resolution imaging[24], and fast live-cell imaging[25]. In our case, we extend this technique into a nonlinear optics regime by including nonlinear excited evanescent illumination, thanks to flexible control over $\mathbf{k}_3$. Using such an FP technique, a complete image of the double slits can be reconstructed by stacking three portions in the k-space along the $k_x$ axis [Fig. 3(c)], which includes two evanescent waves at the frequency $\omega_3$ with opposite directions at $k_{3,\text{eva1}} = 1.56 k_{3,0}$, $k_{3,\text{eva2}} = -1.58 k_{3,0}$, and one near-normal illumination at $k_{3,\text{eva3}} = -0.12 k_{3,0}$. After the iteration converges, an intensity dip appears in the reconstructed image, making the two slits resolvable now [Fig. 3(i)]. Experimentally, a pair of 90nm-width slits spaced 50nm apart on SOI substrate [Fig. 3(j)] can be successfully resolved using this technique as shown in Fig. 3(i), 3(k). The image resolution is sharply improved, in contrast, over those images formed under probe beam normal illumination [Fig. 3(g)] and solely signal illumination at $k_{3,eva1}=1.56k_{3,0}$ [Fig. 3(h)]. these results are also well confirmed by the numerical simulation. Due to the current image system's limitation, e.g. NA, the estimated image resolution is around 50nm (gap distance). To further shrink the gap distance and improve the system resolution, the wave vector $k_{3,eva}$ must be increased accordingly.

At last, we implement this sub-wavelength imaging technique for a nano-slit array with 110nm slit width and 400nm period in Fig. 4. The fine features are clearly revealed [Fig. 4(b-e)] using this technique as compared to the blurred one [Fig. 4(a), 4(c)] under the probe normal illumination. And image contrast has also been significantly improved as shown in Fig. 4(e) because such an FP technique helps to replenish the high spatial frequency parts of images which majorly contribute to a high signal-to-noise ratio (SNR). In the meantime, localized near-field evanescent waves serve as an excellent dark-field illumination source. After filtering out the pump/probe frequencies, they clear out background noises for the high SNR. Moreover, this iterative reconstruction algorithm enables phase-retrieval ability in revealing sensitive information like depth/height[35] and is potential for topographic imaging applications in the future.

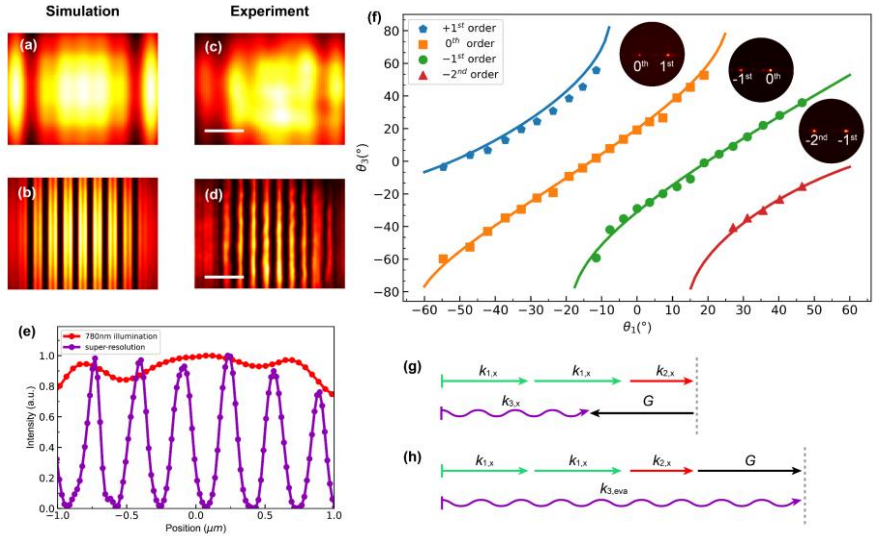

**FIG. 4. Super-resolved nano-slit grating and evanescent wave excitation on the grating.** (a-b) Simulated images of a slit array with 110nm slit width and 400nm period by probe beam illumination and our super-resolution method, respectively (c-d) Experimental results. Scale Bar: $1\mu m$ (e) Cross-section comparison of the two cases, showing great improvement in resolution by our method (f) Experimental demonstration of modified partial phase-matching condition mixed with grating modes when $\theta_2=41°$, solid lines are theoretical prediction calculated from Eq. (4), insets show typical k-space images used to estimate signal's output angles (g) Partial phase-matching condition mixed with grating modes $-G$, making FWM signal become propagating waves (h) Partial phase-matching condition mixed with $G$, resulting in a further increment of $k_{3,eva}$.

Interestingly, this grating structure can also provide a platform facilitating evanescent wave excitation. The grating introduces an additional in-plane momentum $|\mathbf{G}|=|\pm 2\pi/D|$ (first-order), inversely proportional to the spatial period $D$, to FWM processes. These grating modes can be involved in the nonlinear process[12,13] and affect the generated signal beam's wave vector[27]. Under this new scheme, the new phase-matching condition should include the grating momentum term as:

$$k_3 \sin(\theta_3) = 2k_1 \sin(\theta_1) - k_2 \sin(\theta_2) \pm nG \quad (4)$$

Accordingly, Fig. 4(f) demonstrates the angle dependence of FWMs on a grating with a period of 320nm extracted from Fourier space. For a fixed input angle $\theta_2 = 41°$, the measured FWM angles are well-matched with theoretical results from Eq. (4) for both $\pm 1^{st}$ and $-2^{nd}$ grating orders. But $+1^{st}$ grating mode tends to allow larger FWM output angles than the other two orders even for small input $\theta_1$, showing potential for enlarging the magnitude of FWM evanescent wave vector further. This can also be understood by in-plane momentum conservation laws in Fig. 4(g), 4(h), where the additional grating momentum of $+1^{st}$ order further enlarges the FWM wave vector into the evanescent regime. For example, given the pump/probe incident angles at their NA's maximum, the maximum wave vector of the FWM signal can be as large as $4.6k_{3,0}$. This implies the image resolution can be further enhanced if an image object is placed on top of this grating, which requires future investigation.

## 4. Discussions

In the current configuration, the resolution of our imaging technique can reach around 50nm, $\sim \lambda/15.6$ with respect to the probe's wavelength. In the future, a UV laser source with wavelengths smaller than 300nm and the assistance of the aforementioned nano-grating structures may finally put the resolution limit down to a few tens of nanometers, showing great potential in sub-wavelength imaging. Especially, we expect this technique could be beneficial for semiconductor metrology, where UV absorption would not be an issue for semiconductor materials. As for biomedical imaging applications, our technique offers a label-free, scan-free, and far-field super-resolution capability, which is much demanded in this area. To avoid possible laser damage to the bio-imaging samples, we may consider previous similar approaches using nonlinear excited surface plasmon waves to separate laser focus and light-sensitive bio-samples[11]. But compared with fixed wavelength SPP/LSP illumination, our method allows continuous varying of wave vectors, which enables nonlinear FP reconstruction for better imaging qualities. Moreover, given the nonlinear nature of FWM, we also expect our method combined with the coherent anti-stokes Raman scattering (CARS) technique[23,36,37] together may offer chemical-specified, far-field super-resolution imaging, by pairing pump/probe beams' frequencies according to molecules' vibrational energy. At last, in a reversed manner, such excited evanescent waves with large spatial wave vectors are capable of focusing light into tiny spots below the Abbe's diffraction limit in a similar way, this enables a possibility for a new type of high-resolution photolithography mechanism on silicon's surface[38].

## 5. Conclusions

In conclusion, we experimentally realize a super-resolution imaging method based on surface nonlinear FWM excited evanescent wave illumination, which enables label-free, far-field imaging well beyond Abbe's diffraction limit. Such locally excited evanescent waves may also be beneficial for other applications beyond imaging like photolithography.

**Acknowledgment**:
This work was supported by the National key research and development program (Grant No. 2016YFA0302500, 2017YFA0303700); National Natural Science Foundation of China (Grant No. 92050113, No. 11674228). Shanghai MEC Scientific Innovation Program (Grant No. E00075).